\documentclass [12pt]{article}
\usepackage{amssymb,amsmath, amsthm}
\usepackage{graphicx}
\usepackage{xcolor}

\pdfoutput=1 % if your are submitting a pdflatex (i.e. if you have
\usepackage{jheppub} % for details on the use of the package, please
\usepackage{slashed}
\usepackage{color}
\usepackage{xcolor}

\usepackage{graphicx}
\usepackage{caption}
\usepackage{subcaption}

\graphicspath{{Images/}}

\newcommand{\bs}[1]{\boldsymbol{#1}}

\newcommand{\D}{\mathrm{d}}

\title{\boldmath Color coherence in a heavy quark antenna radiating gluons inside a QCD medium}

\author[a,b]{Manoel R. Calvo}
\author[a,c]{Manoel R. Moldes}
\author[a]{Carlos A. Salgado}
\affiliation[a]{Departamento de F\'{i}õsica de Part\'{i}culas and IGFAE, Universidade de Santiago de Compostela, \\15706 Santiago de Compostela, Galicia-Spain}
\affiliation[b]{Institut de Physique Theorique, Saclay}
\affiliation[c]{Centre de Physique Th\'eorique, \'Ecole Polytechnique, CNRS, 91128 Palaiseau, France}

\abstract{We compute the color coherence effects for soft gluon radiation off antennas containing heavy quarks in the presence of a QCD medium --- in color singlet, triplet or octet global states. This work completes the studies of antenna radiation inside a medium which provide a useful picture of the relevance of interference effects in jet parton showers  for the jet quenching phenomenon observed in high-energy nuclear collisions. The analysis is performed resumming the multiple scatterings of the partonic system with the medium. The main conclusion is that decorrelation due to color rotation is more effective in the case in which at least one of the emitters of the antenna is a heavy quark. This effect, present both for a heavy-quark-antiquark  or a heavy-quark-gluon antenna is more relevant for the later or for the case in which the energies of the quark and antiquark are very different. The parameter controlling these  effects involves the dead-cone angle. We find that interferences are cancelled, spoiling the color correlation of the pair, when $\theta_{DC}\equiv M/E\gg 1/\sqrt{\omega L}$ where $E$ and $\omega$ are the energies of the heavy quark and the radiated gluon and $L$ is the medium length. In the case of a heavy-quark-antiquark antenna $t_{\rm form}$ appears instead of $L$ if the original splitting is symmetric. The presence or absence of interferences modifies the energy loss pattern. 
}

\emailAdd{manoel.rodriguez@usc.es}
\emailAdd{manoel.rodriguez-moldes@usc.es}
\emailAdd{carlos.salgado@usc.es}

\begin{document}

\maketitle

\section{Introduction.}
\label{Introduction}

Jets traversing QCD matter created in high-energy nuclear collisions have been experimentally studied for the last ten years, first at the RHIC at BNL \cite{RHIC} and then at the LHC at CERN \cite{Aamodt:2010jd,Aad:2010bu,Chatrchyan:2011}. Despite the success of a theory of jet quenching based on an enhancement of the gluon radiation induced by the medium \cite{Baier:1996kr,Baier:1996sk,Zakharov:1996fv,Zakharov:1997uu,Wiedemann:2000za,gyu00,Wang:2001ifa,Arnold:2002ja} (for a recent review see e.g. \cite{Mehtar-Tani:2013pia}) mainly for the description of the data on inclusive particle suppression, a complete theory, suitable for a consistent and rigorous interpretation of the reconstructed jet data, is still being developed. On the theoretical side, progress in the last few years has been reached in different fronts, as improvements in the splitting probability \cite{Apolinario:2012vy,Blaizot:2012fh}, the use of effective theories as SCET \cite{D'Eramo:2012jh,Fickinger:2013xwa}, or the study of multi-parton radiation \cite{Fickinger:2013xwa}. A systematic program to understand the in-medium intra-jet color coherence effects, using the {\it antenna} setup, has also been started, leading to extremely valuable information about the role of interferences \cite{MehtarTani:2010ma,MehtarTani:2011tz,CasalderreySolana:2011rz,Armesto:2011ir,MehtarTani:2011gf,MehtarTani:2012cy,MehtarTani:2011jw}. A clear new picture of the jet quenching phenomena is emerging from these studies \cite{CasalderreySolana:2012ef,Blaizot:2013vha}.

Since the antenna spectrum plays a central role in understanding jets physics in vacuum \cite{Mueller:1981ex,Ermolaev:1981cm,bas83,Konishi:1979cb,Dokshitzer:1991wu}
, it seems quite natural to ask about the case in what the antenna is traversing a dense medium. The basic question which is addressed with the antenna setup is to which extent subsequent gluon emissions can be considered as independent, hence providing a clear probabilistic picture, and how and when this independency is broken. The well-known results of color coherence in the vacuum lead to the picture of angular-ordered emissions of gluons. The picture in the medium follows the same basic principles and can be simplified as follows \cite{CasalderreySolana:2012ef}: medium-induced radiation can only resolve objects (emitters) which are separated more than a transverse distance $\Lambda_{\rm med}$ determined by the medium properties. Taking the example of a quark-antiquark antenna, when the transverse separation $r_\perp\sim \theta_{q\bar q}L\ll \Lambda_{\rm med}$ the medium cannot resolve the quark and the antiquark individually, so they remain color correlated and emit coherently. This emission can be decomposed into medium-induced (soft and collinear finite) radiation by the total charge, i.e. no radiation if the pair is in a singlet state or radiation as a gluon if the pair is in octet state, plus a vacuum like component which is soft and collinear divergent and angular ordered. In the opposite case $r_\perp\sim \theta_{q\bar q}L\gg \Lambda_{\rm med}$ the medium is very efficient in destroying the color correlation of the pair and the quark and antiquark emit incoherently. Two components can be also distinguished, two medium-induced (soft and collinear finite) contributions each with strength $C_F$ as it corresponds to a singlet emitter, plus a vacuum-like radiation (soft and collinear divergent) but where angular ordered is removed, as expected from two independent color sources.

The mass of the heavy quarks is known to modify the role of color coherence, e.g. introducing a dead-cone angle where radiation is strongly suppressed or removing the strict angular ordering in the vacuum. In the case of the medium,  similar modifications were observed some time ago for the case of a single emitter \cite{Dokshitzer:2001zm,Armesto:2003jh,Zhang:2003wk,Djordjevic:2003zk}: on the one hand, a suppression of the radiation was predicted in most of the phase space relevant for the phenomenological applications; on the other hand, the smaller typical formation time of the gluons produced off massive quarks lead to a reduction of the Landau-Pomeranchuk-Migdal suppression enhancing the radiation at small angles which to some extent {\it fills the dead cone} \cite{Armesto:2003jh}. This additional radiation is not very relevant for the total energy loss of heavy quarks which turns out to be smaller than for light quarks. The corresponding experimental search of this {\it dead-cone effect} lead to one of the still unsolved puzzles in RHIC data \cite{Abelev:2006db,Adare:2010de} in which a suppression of the non-photonic electrons (expected to be dominated by heavy-quark decays) is compatible, taking at face value, with no mass-effect in the radiation. Recent LHC results seem to indicate that indeed heavy-quarks lose less energy than light partons \cite{ALICE:2012ab,Abelev:2012qh,Chatrchyan:2012np} but the actual effect needs to still to be quantified as several different mechanisms contribute to the observed suppression of heavy mesons --- e.g. the different slopes of the perturbative spectra or the harder fragmentation functions in the case of heavy quarks. 

Motivated by these theoretical and experimental findings we present here a calculation of the color coherence effects in a set up which includes a heavy-quark and a gluon antenna radiating a soft gluon in the presence of a medium. The calculation is done resuming the multiple scatterings of the partons involved with the surrounding medium. We also comment on the case of a heavy quark - antiquark antenna, previously studied in Ref. \cite{Armesto:2011ir} but only for the case of the first order in the opacity expansion. The main result in the paper is the reduction of the color coherence when the dead-cone angle $\theta_{DC}\equiv M/E\gg 1/\sqrt{\omega L}$ where $E$ and $\omega$ are the energies of the heavy quark and the radiated gluon and $L$ is the medium length\footnote{Notice that in this paper,  light-cone coordinates are used, so, the definition of the dead-cone angle $\theta_{DC}$ will be slightly different, see below.}. This suppression of the color coherence enhances the phase space for radiation, hence the energy loss of the heavy quark. Whether this enhancement can lead to a sizable suppression of the heavy quarks in the experimental environment is left for a future study. The present publication completes the study of the in-medium antenna radiation spectrum \cite{MehtarTani:2010ma,MehtarTani:2011tz,CasalderreySolana:2011rz,Armesto:2011ir,MehtarTani:2011gf,MehtarTani:2012cy,MehtarTani:2011jw} and provides the way to encode heavy quark effects in the new picture of jet quenching being developed.

\section{Amplitudes and formalism.} 

The derivation of the spectrum is similar to the ones presented in previous works \cite{MehtarTani:2010ma,MehtarTani:2011tz,CasalderreySolana:2011rz,Armesto:2011ir,MehtarTani:2011gf,MehtarTani:2012cy,MehtarTani:2011jw} but including the mass of the heavy quark explicitly. 
The amplitude for one gluon emission is calculated using the reduction formula 

\begin{equation}
\label{amplitude}
\mathcal{M}^{a} (k)=-\sum\limits_{\lambda}\int\limits_{x^{+}=+\infty} \D x^- \D^2\mathbf{x} \,\mbox{e}^{ik \cdot x} \,2 \partial^+_x \mathbf{A}^{a}(x)\cdot \mbox{\boldmath{$\epsilon$}}_{\lambda}(\vec{k})
\end{equation}
with $k^\mu=(\omega,\vec{k})$ being the 4-momentum of the emitted gluon and $\mathbf{A}$ the transverse gauge field.
The gauge field is obtained from the classical Yang-Mills (CYM) equations
\begin{equation}
[D_{\mu}, F^{\mu\nu}]=J^{\nu} \\
\end{equation}
where $D_\mu\equiv\partial_\mu-igA_\mu$ and $F_{\mu\nu}\equiv\partial_\mu A_\nu-\partial_\nu A_\mu-ig[A_\mu,A_\nu]$, and with the current $J^\mu$ being covariantly conserved, i.e., 
\begin{equation}
\lbrack D_{\mu}, J^{\mu} \rbrack=0
\end{equation}

The current $J^\mu$ has three components: one of them for each leg of the antenna and a third one representing the highly virtual particle coming from a hard process that splits into the antenna pair.  For the case of a $Qg$ antenna, this means that $J^\mu=J_q^\mu + J_g^\mu+J_3^\mu$, being the three components are the currents representing the quark, the gluon and the virtual quark, respectively.

The initial state of the antenna is given by the vacuum current $J_{(0)}^\mu=J_{q(0)}^\mu + J_{g(0)}^\mu+J_{3(0)}^\mu$, where
\begin{equation}
\label{vaccumcurrent}
J^{\mu,a}_{i(0)} (x)=g \frac {p_i^{\mu}}{E_i} \delta^{(3)}\left( \vec {x} -\frac{\vec{p_i}}{E_i} t\right) \theta(t) Q_i^a
\end{equation}
represents a particle with momentum $p_i^\mu=(E_i,\vec{p_i})$ and charge color vector $Q_i^a$ and $i=q,g,3$.

The $J_3^\mu$ current is needed for charge and mometum conservation ($Q_3=Q_q+Q_g$ and $\vec{p_3}=-\vec{p_q}-\vec{p_g}$, respectively), but in a colored antenna it does not contribute  in the frame where $p_3^\mu \simeq(0, p_3^- , \bs 0)$\footnote{Any 4-vector $a^\mu=(a^0,a^1,a^2,a^3)$ is expressed in light-cone coordinates as $a^\mu=(a^+,a^-,\mathbf{a})$, with $a^{\pm}=(a^0\pm a^3)/\sqrt{2}$ and $\mathbf{a}=(a^1,a^2)$} due to the light-cone gauge we perform our calculation in ($A^+=0$). In the case of a singlet antenna, $J_3^\mu$ does not contribute because $Q_3=0$.

%The initial state of a component of the antenna with momentum $p^\mu=(E,\vec{p})$ and charge color vector $Q^a$ is represented by this vacuum current:
%\begin{equation}
%\label{vaccumcurrent}
%J^{\mu,a}_{(0)} (x)=g \frac {p^{\mu}}{E} \delta^{(3)}\left( \vec {x} -\frac{\vec{p}}{E} t\right) \theta(t) Q^a
%\end{equation}

The effect of the medium over the vacuum current $J_{(0)}^\mu$ is to induce a color rotation:
\begin{equation}
J^{\mu}(x)=U_p(x^+,0)J^{\mu}_{q(0)}(x)+U_{\bar p}(x^+,0)J^{\mu}_{g(0)}(x)
\end{equation}
described by a Wilson line:
\begin{equation}
U_p(x^+,0;\mathbf{r})\equiv \mathcal{P}\, \mbox{exp}\, \left\{\int_0^{x^+}\D \xi\, T\cdot A_{\mathrm{med}}^-(\xi, \mathbf{p}\xi/p^+)\right\}
\end{equation}
where we have denoted the quark momentum as $p$ and the gluon mometum as $\bar p$.

Leaving only terms linear on the medium induced field %and performing the calculation in the light-cone gauge ($A^+=0$)\footnote{Any 4-vector $x^\mu=(x^0,x^1,x^2,x^3)$ is expressed in light-cone coordinates as $x^\mu=(x^+,x^-,\mathbf{x})$, with $x^{\pm}=(x^0\pm x^3)/\sqrt{2}$ and $\mathbf{x}=(x^1,x^2)$},
and focusing on the quark current (the calculation with the gluon current is the same and so it gives an analogous result) we get the following expression for the amplitude for emission off the quark:

\begin{equation}
\begin{split}
\label{amplitude2}
\mathcal{M}_q^a(\vec{k})&=\sum\limits_\lambda\frac{g}{k^+} \int\limits_{x^+=+\infty} \D^2\mathbf{x}\,e^{ik^-x^+} e^{-i\mathbf{k}\cdot \mathbf{x}}  \int_0^{+\infty} \D y^+\, e^{ik^+ \frac{p^-}{p^+} y^+}\\
&\times\mbox{\boldmath{$\epsilon$}}_\lambda(k) \cdot (i \boldsymbol{\partial} _{y} + k^+\mathbf{n})\, \mathcal{G}^{ab} (x^+,\mathbf{x}; y^+,\mathbf{y}\vert k^+)\Big\vert_{\mathbf{y}=\mathbf{n}y^+}U^{bc}_p(y^+, 0)Q_q^c
\end{split}
\end{equation}
where we have explicited the color structure, defined the dimensionless vector $\mathbf{n}=\mathbf{p}/p^+$ and $\mathcal{G}$ is a Green's function that takes into account both the color rotation of the gluon and its Brownian motion in the transverse plane due to interactions with the medium field. These features of the Green's function $\mathcal{G}$ can be easily seen thanks to its expression as a path integral in the transverse plane: 
\begin{equation}
\mathcal{G}(x^+, \mathbf{x};y^+,\mathbf{y}\vert k^+)=\int\limits_{\mathbf{r}(y^+)=\mathbf{y}}^{\mathbf{r}(x^+)=\mathbf{x}} \mathcal{D}\mathbf{r}\, \mbox{exp}\left\{\frac{ik^+}{2}\int_{y^+}^{x^+}\D\xi\, \dot{\mathbf{r}}^2(\xi)\right\}U(x^+,y^+;\mathbf{r})
\end{equation}

The mass effects can be now easily identified, as they enter the $-$-component of the 4-momentum heavy-quark  momentum through the dispersion relation
%
%Reached this point, we wonder what could be the effect due to the presence of a mass $M$. To see that, we recall that the only difference between a massive quark and a massless one comes from the dispersion relation:
\begin{equation}
 2p^+p^- - \mathbf{p}^2= M^2.
\end{equation}
Taking the eikonal limit, in which the quark follows a straight line in the direction ${\bf n}$, the presence of the mass appears only in the phase
%
%So we can understand the existence of a mass as a modification to the relation of  $p^-$ with respect of the other components of the momentum. Since the only place where this $p^-$ component appears (provided we neglect any dependence of the medium field on the $x^-$ variable) in the previous expressions is in an exponential, the presence of mass brings us a new phase:
%
\begin{equation}
\mbox{exp}\left(i k^+\frac{p^-}{p^+} y^+\right) =\mbox{exp}\left(i \frac{k^+}{2}\theta^2_{DC}y^+\right)\,\mbox{exp}\left(i \frac{k^+}{2}\mathbf{n}^2y^+\right)
\end{equation}
where $\theta_{DC}$ is the so called \textit{dead-cone angle}, defined as\footnote{Notice that this definition is slightly different that the one given in section \ref{Introduction}. Translating the light-cone coordinates into space ones, $p^+\simeq \sqrt{2}E$, the dead-cone angle reads $\theta_{DC}\simeq M/(\sqrt{2}E)$}
\begin{equation}
\label{deadcone}
\theta_{DC}\equiv \frac{M}{p^+}
\end{equation}
%
%From this point on, we denote with a bar quantities (such as momentum) related to the gluon leg of the antenna, while those related to the quark leg are left without bar; e.g., $p^\mu$ represents the quark momentum and $\bar{p}^\mu$ represents the gluon momentum.
%
We have now all the ingredients to explicitly write the two contributions to the amplitude; when the gluon is radiated off the heavy quark
\begin{equation}
\begin{split}
\label{amplitude_q}
\mathcal{M}_q^a(\vec{k})&=\sum\limits_\lambda\frac{g}{k^+} \int\limits_{x^+=+\infty} \D^2\mathbf{x}\,e^{ik^-x^+} e^{-i\mathbf{k}\cdot \mathbf{x}}  \int_0^{+\infty} \D y^+\, \mbox{exp}\left[i\frac{k^+}{2}(\theta_{DC}^2+\mathbf{n}^2)y^+\right]\\
&\times\mbox{\boldmath{$\epsilon$}}_\lambda(k) \cdot (i \boldsymbol{\partial} _{y} + k^+\mathbf{n})\, \mathcal{G}^{ab} (x^+,\mathbf{x}; y^+,\mathbf{y}\vert k^+)\Big\vert_{\mathbf{y}=\mathbf{n}y^+}U^{bc}_p(y^+, 0)Q^c_q
\end{split}
\end{equation}

\noindent 
and when the gluon is radiated off the gluon
\begin{equation}
\begin{split}
\label{amplitude_g}
\mathcal{M}_g^a(\vec{k})&=\sum\limits_\lambda\frac{g}{k^+} \int\limits_{x^+=+\infty} \D^2\mathbf{x}\,e^{ik^-x^+} e^{-i\mathbf{k}\cdot \mathbf{x}}  \int_0^{+\infty} \D y^+\, \mbox{exp}\left[i\frac{k^+}{2}\bar{\mathbf{n}}^2 y^+\right]\\
&\times\mbox{\boldmath{$\epsilon$}}_\lambda(k) \cdot (i \boldsymbol{\partial} _{y} + k^+\bar{\mathbf{n}})\, \mathcal{G}^{ab} (x^+,\mathbf{x}; y^+,\mathbf{y}\vert k^+)\Big\vert_{\mathbf{y}=\bar{\mathbf{n}}y^+}U^{bc}_{\bar{p}}(y^+, 0)Q^c_g
\end{split}
\end{equation}

We can explicitly see the difference due to the presence of mass: a complex phase in \eqref{amplitude_q} that \eqref{amplitude_g} lacks.

%%%%%%%%%%%%% SECTION

\section{Radiation Spectrum of the Heavy Quark-Gluon Antenna.}

With the amplitudes \eqref{amplitude_q} and \eqref{amplitude_g} we can compute the radiation spectrum of the antenna. A particularly convenient way of presenting this spectrum is by separating the independent radiation (in the vacuum easily identified by the two collinear divergencies) of the two emitters

\begin{equation}
dN=\frac{\alpha_s}{(2\pi)^2} \left[ C_F \mathcal{R}_{q}+ C_A \mathcal{R}_{g} -C_A \mathcal{J}\right] \frac{d^3k}{(k^+)^3} 
\end{equation}

\noindent
Here, we have defined the independent radiation off the heavy quark $\mathcal{R}_q$ by

\begin{equation}
\label{R_q}
C_F \mathcal{R}_q=(k^+)^2 \langle |\mathcal{M}_q|^2 \rangle
\end{equation}

\noindent
the independent radiation off the gluon $\mathcal{R}_g$

\begin{equation}
\label{R_g}
C_A \mathcal{R}_g=(k^+)^2 \langle |\mathcal{M}_g|^2\rangle
\end{equation}

\noindent
and the interference spectrum between both emitters $\mathcal{J}$

\begin{equation}
-C_A\mathcal{J}= (k^+)^2\, \mbox{Re}\langle \mathcal{M}_q \mathcal{M}_g^\dagger \rangle
\end{equation}

%\subsection{Independent Radiation Spectrum.}

\noindent
The spectrum of independent radiation off the heavy quark $\mathcal{R}_q$ is directly evaluated from \eqref{R_q} taking the limit $\mathbf{n}\rightarrow 0$ 

\begin{multline}
\label{heavy R_q}
\mathcal{R}_q=2\mathrm{Re} \int_0^\infty \!\D y'^+ \int_0^{y'^+}\!\D y^+ \, \mbox{exp}\left[i\frac{k^+}{2}\theta_{DC}^2(y^+-y'^+)\right] \times\\
\times \int\D^2\mathbf{z}\,\, \mathrm{exp}\left[-i\mathbf{k}\cdot\mathbf{z}-\frac{1}{2}\int_{y'^+}^\infty\D\xi\, n(\xi)\sigma(\mathbf{z})\right]\, \boldsymbol{\partial}_y \cdot \boldsymbol{\partial}_{z}\, \mathcal{K}(y'^+,\mathbf{z};y^+,\mathbf{y})\big\vert_{\mathbf{y}=\mathbf{0}}
\end{multline}

\noindent which, as expected, is exactly the same expression derived previously for the medium-induced gluon radiation off a single heavy quark in the BDMPS multiple scattering approximation \cite{Armesto:2003jh}. This confirmation is also a test of our formalism.

In the same manner, the medium-induced gluon radiation off a gluon is obtained from \eqref{R_g}
\begin{multline}
\mathcal{R}_g=2\mathrm{Re} \int_0^\infty \!\D y'^+ \int_0^{y'^+}\!\D y^+ \int\D^2\mathbf{z}\,\, \mathrm{exp}\left[-i\mathbf{k}\cdot\mathbf{z}-\frac{1}{2}\int_{y'^+}^\infty\D\xi\, n(\xi)\sigma(\mathbf{z})\right]\times\\
\times \boldsymbol{\partial}_y \cdot \boldsymbol{\partial}_{z}\, \mathcal{K}(y'^+,\mathbf{z};y^+,\mathbf{y})\big\vert_{\mathbf{y}=\mathbf{0}}
\end{multline}
which again coincides with the known results.

%\subsection{Interference Spectrum.}

The most interesting part of our analysis is, of course, the color coherent emission off the two emitters given by the interference term $\mathcal{J}$ 

\begin{equation}
\begin{split}
\label{J}
\mathcal{J} = \mathrm{Re}\Bigg\{&\int_0^{\infty}\D y'^+ \int_0^{y'^+}\D y^+\, \left(1- \Delta_{\rm med} (y^+,0)\right) \,\mbox{exp}\left[i\frac{k^+}{2}(\theta_{DC}^2+\delta\mathbf{n}^2)y^+\right]\times\\
&\times \int\D^2\mathbf{z}\,  \mbox{exp} \left[-i\bar{\boldsymbol{\kappa}}\cdot \mathbf{z} - \frac{1}{2} \int_{y'^+}^\infty\D\xi\, n(\xi ) \sigma(\mathbf{z})\right]\times\\
&\times(\boldsymbol{\partial} _y -ik^+\delta\mathbf{n}) \cdot \boldsymbol{\partial} _z\, \mathcal{K}(y'^+,\mathbf{z};y^+,\mathbf{y})\big\vert_{\mathbf{y}=\delta\mathbf{n}y^+}\Bigg\}+\mbox{sym.}
\end{split}
\end{equation}
with $\bar{\boldsymbol{\kappa}}=\mathbf{k}-k^+\bar{\mathbf{n}}$ (and $\boldsymbol{\kappa}=\mathbf{k}-k^+\mathbf{n}$), $\delta\mathbf{n}=\mathbf{n}-\mathbf{\bar{n}}$ and $\mathcal{K}$ being the path integral
\begin{equation}
\mathcal{K}(y'^+,\mathbf{z};y^+,\mathbf{y}\vert k^+)=\int\limits_{\mathbf{r}(y^+)=\mathbf{y}}^{\mathbf{r}(y'^+)=\mathbf{z}}\mathcal{D}\mathbf{r}\, \mbox{exp}\left\{\int_{y^+}^{y'^+}\D\xi\, \left(i\frac{k^+}{2}\dot{\mathbf{r}}^2(\xi)-\frac{1}{2}n(\xi)\sigma(\mathbf{r})\right)\right\}
\end{equation}
that takes into account the emitted gluon multiple scattering with the medium and its Brownian motion in the transverse plane from $\mathbf{r}(y^+)=\mathbf{y}$ to $\mathbf{r}(y'^+)=\mathbf{z}$. The symmetric part is obtained exchanging $q\leftrightarrow g$.

Equation (\ref{J}) is the main result of this paper. Compared with the light-quark case, the only difference is the presence of a new phase including the dead-cone angle, $\theta_{\rm DC}$. This is similar to the case found for the single heavy quark emitter  \cite{Armesto:2003jh} and a direct consequence of the eikonal approximation assumed in the calculation. The role of the phase is to suppress the radiation in some regions of phase space due to the mass effects. 

\section{Discussion and conclusions.}

We will focus here in the two main terms which suppress the interferences, Eq. (\ref{J}), namely the decoherence parameter $\Delta_{\rm med}$
\begin{equation}
\label{eq:deltamed}
\Delta_{\rm med}(y^+,0)\equiv 1-\mbox{exp}\left[-\frac{1}{2}\int_0^{y^+}\D\xi\, n(\xi)\sigma(\delta\mathbf{n}\, \xi)\right]\simeq
1-\mbox{exp}\left[-\frac{1}{12}\hat q\, \delta{\bf n}^2 L^3\right]
\end{equation}
(where we have used the multiple soft scattering approximation for the last expression, $\hat q$ being the transport coefficient also known as {\it jet quenching parameter}) and the {\it dead-cone} phase 
\begin{equation}
\label{eq:deadconephase}
\Delta_{DC}(y^+,y'^+)=\exp\left[i\frac{k^+}{2}\left(\theta_{DC}^2\, y^+-\bar\theta_{DC}^2 \,y'^+\right)\right]
\end{equation}
Notice that for a (triplet) $Qg$ antenna,  $\bar\theta_{DC}=0$ and we recover the phase in Eq. (\ref{J}). For an octet, or singlet, $Q\bar Q$, there are two dead-cone angles, each one referring to the energy of the corresponding emitter, $\theta_{DC}=M/p^+$ and $\bar\theta_{DC}=M/\bar p^+$. Notice also that in the $Q\bar{Q}$ case, when one of the emitters is much more energetic than the other, say $\bar p^+\gg p^+$, i.e. for very asymmetric $g\to Q\bar Q$ splittings, the suppression pattern of the interferences is similar to that of the $Qg$ case, as  $\bar \theta_{DC}\ll \theta_{DC}$ in (\ref{eq:deadconephase}).

The effect of the decoherence parameter $\Delta_{\rm med}$ in the in-medium antenna radiation has been discussed at length in the previous calculations, in particular in Ref. \cite{CasalderreySolana:2012ef}. Its role is to suppress the interference terms (\ref{J}) when the transverse size of the antenna is larger than the typical medium color correlation length in the transverse plane. I.e. when the correlation length is larger than the size of the pair, $\Delta_{\rm med}\to 0$, the medium cannot resolve the individual emitters, which act as a single object with the total charge of the pair ($C_F$ for triplet, $C_A$ for octet or $0$ for singlet). In the opposite case, $\Delta_{\rm med}\to 1$, the medium resolves the antenna and breaks the color coherence of the pair so that they behave as two independent particles.

Consider, for simplicity the case of a color octet $Q\bar Q$ antenna in the symmetric case $(p^+\simeq p'^+)$ and assuming that the opening angle $\delta{\bf n}^2\sim \theta_{q\bar q}^2$ is small (which is a comfortable assumption $\theta_{q\bar q}^2\ll k^+L$) and can be neglected in the phase. When color decoherence happens, $\Delta_{\rm med}\sim 1$, the radiation is that of two independent emitters 

\begin{equation}
\label{eq:cotetincoh}
(k^+)^3\frac{dN}{d^3k}=\frac{\alpha_s}{(2\pi)^2} C_F \left(\mathcal{R}_q+\mathcal{R}_{\bar q}\right)
\end{equation}
and the spectrum is just the superposition of two spectra for gluon radiation off a heavy-quark. This behavior was first identified for the massless case in \cite{MehtarTani:2011tz}. 

An interesting case appears in the totally coherent case, $\Delta_{\rm med}\sim 0$, in which the medium cannot resolve the pair, and one has ${\cal R}_q\sim{\cal R}_{\bar q}\sim{\cal J}$. Now,  the medium-induced radiation is simply

\begin{equation}
(k^+)^3\frac{dN}{d^3k}=\frac{\alpha_s}{(2\pi)^2} C_A \mathcal{R}_q
\end{equation}
That would correspond to the radiation {\bf off a gluon} (as the total charge gives the factor $C_A$) {\bf but with the dead-cone suppression factor}. The production of heavy quarks is dominated by gluon splitting at large transverse momenta. This tells that the corresponding energy loss will be as the one for gluons but with a dead-cone suppression as long as the pair stays in a color coherent state. The corresponding time for this to occur can be easily estimated from Eq. (\ref{eq:deltamed}) to be $t_{\rm coh}\sim[12/(\theta_{q\bar q}^2\hat q)]^{1/3}$ and can be sizable according to the estimates in \cite{CasalderreySolana:2012ef}. Notice that in this case, the total radiation is still larger than the sum of the radiation of the $Q\bar Q$ pair (\ref{eq:cotetincoh}) as $2C_F<C_A$. (For the rest of the color configurations, color coherence reduces the amount of energy loss). 

Another consequence of our analysis is that for the case of asymmetric splitting, either a $Qg$ antenna or a $Q\bar Q$ antenna with very different energies of quark and antiquark, the suppression factor is much larger than for the symmetric case. Indeed, in this last case, the  suppression will happen for gluon energies $\omega > 1/ (t_{\rm form}\theta_{DC}^2)$, while it will happen much earlier, $\omega > 1/ (L\theta_{DC}^2)$ for the asymmetric configuration.

All these features provide a nice generalization of the findings of the antenna for the massless case with physics dominated by color coherence effects or the lost of them due to the interaction with the medium. The phenomenological consequences for the heavy quark case will be presented in a separate work.

\section*{Acknowledgments}
This work is supported by European Research Council grant HotLHC ERC-2011-StG-279579, by Xunta de Galicia (Conseller\'\i a de Educaci\'on); by the Spanish Consolider-Ingenio 2010 Programme CPAN and by FEDER. Manoel R. Calvo thanks Ministerio de Ciencia e Innovaci\'on of Spain for financial support under  grant FPI-BES-2010-040744.

\end{document}